\documentclass[final,5p,times]{elsarticle}

\usepackage{booktabs}
\usepackage{tabularx}
\usepackage{textcomp, gensymb}
\usepackage{standalone}

\usepackage{amsmath}

\usepackage{hyperref}

\usepackage{comment}

\usepackage{multirow}
\usepackage{xurl}

\usepackage{amsmath}
\usepackage{amssymb}
\usepackage{float}

\usepackage{color}

\usepackage{algorithm}
\usepackage{algpseudocode}
\usepackage{multicol}

\floatstyle{plain}
\newfloat{myalgo}{tbhp}{mya}

{\begin{myalgo}[#1]
\centering
\begin{minipage}{\textwidth}
\begin{algorithm}[H]}%
{\end{algorithm}
\end{minipage}
\end{myalgo}}

\usepackage{breqn}

\usepackage{lscape}
\usepackage{rotating}

\usepackage{tabularx}

\usepackage{amsmath}
\usepackage{graphicx}
\usepackage{xcolor}
\usepackage{soul}
\usepackage[normalem]{ulem}

\graphicspath{ {./} }

\usepackage{caption}
\definecolor{cadetblue}{rgb}{0.37, 0.62, 0.63}
\definecolor{fuzzywuzzy}{rgb}{0.8, 0.4, 0.4}
\definecolor{babyblue}{rgb}{0.54, 0.81, 0.94}

\usepackage{tikz}
\usepackage{pgfplots}
\usepackage{pgf-umlcd}
\pgfplotsset{compat=1.18} 
\usetikzlibrary{external, shapes}
\usetikzlibrary{calc, matrix, positioning, chains, automata, quotes}
\usetikzlibrary{arrows.meta, backgrounds, ext.positioning-plus, ext.node-families.shapes.geometric, calc}
\usetikzlibrary{decorations.markings}
\usetikzlibrary{decorations.pathreplacing,calligraphy}

\tikzset{
  frame/.style={
    rectangle, draw,
    text centered,
    text width=6em, 
    minimum height=4em,fill=white,
    rounded corners,
  },
  initialize/.style={
    rectangle, draw,
    text centered,
    text width=16em, 
    minimum height=4em,fill=white,
    rounded corners,
  },
  block/.style={
    draw, rectangle, 
    text centered,
    fill=white,
    minimum height=3em, 
    minimum width=5em,
    rounded corners,
  },
  sum/.style={
    draw, circle,
    fill=white,
    node distance=1cm,
  },
  input/.style={coordinate},
  output/.style={coordinate},
  line/.style={ 
    draw, rounded corners=3mm,
  },
  line-rounded/.style={
    draw, -{Latex},rounded corners=3mm,
  },
}

\usepackage{subcaption} 

\usepackage[noabbrev]{cleveref}

\usepackage{amssymb}

\journal{Expert Systems With Applications}

\begin{document}

\begin{frontmatter}

\title{Application of Soft Actor-Critic Algorithms in Optimizing Wastewater Treatment with Time Delays Integration}

\author[inst1,inst2]{Esmaeel Mohammadi\corref{cor1}}
\ead{esm@kruger.dk}
\cortext[cor1]{Corresponding author.}
\author[inst3]{Daniel Ortiz-Arroyo}\ead{doa@energy.aau.dk}
\author[inst1]{Aviaja Anna Hansen}\ead{avh@kruger.dk}
\author[inst1]{Mikkel Stokholm-Bjerregaard}\ead{mxs@kruger.dk}
\author[inst4]{Sébastien Gros}\ead{sebastien.gros@ntnu.no}
\author[inst4]{Akhil S Anand}\ead{akhil.s.anand@ntnu.no}
\author[inst3]{Petar Durdevic}
\ead{pdl@energy.aau.dk}

\affiliation[inst1]{organization={Krüger A/S},
            addressline={Indkildevej 6C}, 
            city={Aalborg},
            postcode={9210}, 
            state={North Jutland},
            country={Denmark}}

\affiliation[inst2]{organization={Department of Chemistry and Bioscience, Aalborg University},
            addressline={Fredrik Bajers Vej 7H}, 
            city={Aalborg},
            postcode={9220}, 
            state={North Jutland},
            country={Denmark}}

\affiliation[inst3]{organization={AAU Energy, Aalborg University},
            addressline={Niels Bohrs vej 8}, 
            city={Esbjerg},
            postcode={6700}, 
            state={South Jutland},
            country={Denmark}}

\affiliation[inst4]{organization={Department of Engineering Cybernetics, Norwegian University of Science and Technology},
            addressline={Høgskoleringen 1}, 
            city={Trondheim},
            postcode={7034}, 
            state={Trøndelag},
            country={Norway}}

\begin{abstract}
Wastewater treatment plants face unique challenges for process control due to their complex dynamics, slow time constants, and stochastic delays in observations and actions. These characteristics make conventional control methods, such as Proportional-Integral-Derivative controllers, suboptimal for achieving efficient phosphorus removal, a critical component of wastewater treatment to ensure environmental sustainability. This study addresses these challenges using a novel deep reinforcement learning approach based on the Soft Actor-Critic algorithm, integrated with a custom simulator designed to model the delayed feedback inherent in wastewater treatment plants. The simulator incorporates Long Short-Term Memory networks for accurate multi-step state predictions, enabling realistic training scenarios. To account for the stochastic nature of delays, agents were trained under three delay scenarios: no delay, constant delay, and random delay. The results demonstrate that incorporating random delays into the reinforcement learning framework significantly improves phosphorus removal efficiency while reducing operational costs. Specifically, the delay-aware agent achieved \textbf{36}\% reduction in phosphorus emissions, \textbf{55}\% higher reward, \textbf{77}\% lower target deviation from the regulatory limit, and \textbf{9}\% lower total costs than traditional control methods in the simulated environment. These findings underscore the potential of reinforcement learning to overcome the limitations of conventional control strategies in wastewater treatment, providing an adaptive and cost-effective solution for phosphorus removal.
\end{abstract}


\end{frontmatter}


\section{Introduction}
Wastewater treatment is a critical process in maintaining environmental sustainability and public health. Traditional control methods often struggle with the complex, nonlinear, and time-varying nature of wastewater treatment plants (WWTP). Reinforcement learning (RL) offers a promising alternative: learning control policies through interaction with the environment, thus adapting to the inherent complexities and uncertainties of the process. Recent studies have demonstrated the potential of RL in optimizing energy consumption, enhancing treatment efficiency, and ensuring regulatory compliance in WWTPs \cite{chen2021optimal, syafiie2011model, croll2023reinforcement, yang2021reinforcement}.

Soft Actor-Critic (SAC) is a state-of-the-art deep reinforcement learning (DRL) algorithm known for its exceptional sample efficiency and stability in continuous action spaces \cite{haarnoja2018soft}. Traditional policy gradient methods, such as Trust Region Policy Optimization (TRPO) \cite{schulman2015trust} and Proximal Policy Optimization (PPO) \cite{schulman2017proximal}, have advanced DRL by providing stable on-policy training. However, they generally require more samples, as they cannot reuse past experiences efficiently \cite{lillicrap2015continuous}. In contrast, SAC is an off-policy algorithm, allowing it to reuse past experiences from a replay buffer, significantly enhancing sample efficiency. SAC effectively balances exploration and exploitation by leveraging a maximum entropy framework, making it especially valuable for complex industrial applications with high exploration costs \cite{nian2020review, haarnoja2018soft}. Its ability to handle high-dimensional state and action spaces enables SAC to develop sophisticated control strategies, a crucial feature for wastewater treatment plants, where multiple interacting variables must be managed concurrently \cite{croll2023reinforcement}.

One significant challenge in applying RL to industrial processes, including WWTPs, is the presence of time delays. Time-delayed systems are prevalent in many industrial settings, and ignoring these delays can lead to suboptimal or unstable control policies \citep{bouteiller2020reinforcement}. Incorporating time delays into the RL framework requires advanced methods to predict future states or adapt policies for delayed feedback. Research has shown that DRL algorithms, when appropriately modified, can effectively manage time delays and improve control performance in such systems \cite{bouteiller2020reinforcement}.

This paper aims to explore the application of the SAC algorithm in the context of WWTP control and optimization, addressing the challenges posed by time delays. Control policies were trained on a state-of-the-art simulator employing Long Short-Term Memory (\emph{LSTM}) networks, which were previously trained and enhanced for multi-step simulations in prior studies \cite{mohammadi2024deep, mohammadi2024improved, mohammadi2024multi}. Through extensive simulations and case studies, we demonstrate the efficacy of SAC in enhancing the operational efficiency and reliability of wastewater treatment processes. Three scenarios and policy algorithms were investigated, considering no delays, constant delays, or random delays in the observations and actions of the system. The results indicated that agents accounting for constant or random delays outperformed those with no delay in cumulative rewards. Furthermore, when random delays were introduced, the agent achieved better performance than with constant delays, highlighting the importance of capturing the stochastic nature of delays in environments such as WWTPs. The findings contribute to the growing knowledge achieved by applying advanced DRL methods in industrial process control, offering insights into future research directions and practical implementations.

\section{Literature Review}
Deep reinforcement learning has promising potential in the process and industrial control field as it provides sophisticated methods for handling complex, nonlinear, and high-dimensional systems. DRL algorithms like Deep Deterministic Policy Gradient (DDPG) and Proximal Policy Optimization have successfully optimized control strategies in chemical processes, enhancing efficiency and adaptability \cite{spielberg2017deep, byun2020robust}. For instance, DRL has been utilized to develop control policies that can handle time-varying dynamics and uncertainties in industrial environments, outperforming traditional control methods that rely on linear approximations or require extensive manual tuning \cite{bao2021deep}.
\begin{table*}[h]
\caption{Comparison of Deep Reinforcement Learning Algorithms for Continuous Control.}
\centering
\label{tab:drl_algorithms}
\begin{tabularx}{\textwidth}{@{}p{2cm}|X|X|X@{}}
\toprule
\textbf{Algorithm} & \textbf{Concept} & \textbf{Advantages} & \textbf{Disadvantages} \\ \midrule
\textbf{DDPG} &
Deterministic policy gradient method with actor-critic architecture; uses target networks and experience replay &
Handles high-dimensional continuous action spaces; sample efficient due to off-policy learning &
Sensitive to hyperparameters and exploration noise; prone to overestimation bias; less robust \\ \midrule
\textbf{Actor-Critic} &
Combines actor (policy) and critic (value function) networks to reduce variance in policy updates &
Balances the benefits of policy-based and value-based methods; faster convergence; improved stability &
Complex implementation; requires careful tuning of actor and critic learning rates; moderate sample efficiency \\ \midrule
\textbf{SAC} &
Entropy-regularized off-policy actor-critic method with stochastic policies &
Encourages exploration via entropy regularization; highly stable and sample efficient; handles stochastic environments well &
Computationally intensive; requires careful tuning of entropy coefficient \\ \midrule
\textbf{PPO} &
Policy gradient method with a clipped objective to ensure stable updates &
Stable and reliable updates; balances simplicity and performance; widely used in continuous control tasks &
Medium sample efficiency; requires careful hyperparameter tuning; not as sample efficient as off-policy methods \\ \bottomrule
\end{tabularx}
\end{table*}


DRL has shown promise in energy management and robotics applications in industrial optimization, specifically in wastewater treatment processes. Studies have demonstrated that DRL can optimize aeration control in wastewater treatment plants, leading to significant energy savings while maintaining effluent quality standards \cite{chen2021optimal}. Despite these successes, challenges such as sample efficiency, safety concerns, and the need for large data for training persist. Ongoing research addresses these issues by developing more efficient algorithms and integrating DRL with model-based approaches to enhance real-world applicability \cite{nian2020review}. Some of the most used DRL algorithms for continuous control with their advantages and disadvantages are summarized in Table \ref{tab:drl_algorithms}.

\subsection{Time Delays}
\colorlet{water}{blue!70!cyan!70}
\colorlet{inside}{orange!50}

\tikzset{
	->,  
	>=stealth, 
	shorten >=2pt, shorten <=2pt, 
	node distance=1.8cm, 
	every state/.style={draw,
 very thick,
 minimum size=0.5cm, 
 inner sep=2pt}, 
	initial text=$ $, 
 }
 
\tikzset{
    table/.style={
        matrix of nodes,
        row sep=-\pgflinewidth,
        column sep=-\pgflinewidth,
        nodes={
            rectangle,
            align=center
        },
        minimum height=1.5em,
        text depth=0.5ex,
        text height=2ex,
        nodes in empty cells,
        every even row/.style={
        },
        column 1/.style={
            nodes={text width=1em,font=\bfseries}
        },
        row 1/.style={
            nodes={
                text=white,
                font=\bfseries
            }
        }
    }
}

\def\SPSB#1#2{\rlap{\textsuperscript{\textcolor{red}{#1}}}\SB{#2}}

\begin{figure*}
\centering
\begin{tikzpicture}
    \node[state with output] (s0) {$\mathbf{s}_0$ \nodepart{lower} $a_0$};
    \node[state with output, right of=s0] (s1) {$\mathbf{s}_1$ \nodepart{lower} $a_1$};
    \node[state with output, right of=s1] (s2) {$\mathbf{s}_2$ \nodepart{lower} $a_2$};
    \node[state with output, right of=s2] (s3) {$\mathbf{s}_3$ \nodepart{lower} $a_3$};
    \node[state with output, right of=s3] (s4) {$\mathbf{s}_4$ \nodepart{lower} $a_4$};
    \node[state with output, right of=s4] (s5) {$\mathbf{s}_5$ \nodepart{lower} $a_5$};
    \node[state with output, right of=s5] (s6) {$\mathbf{s}_6$ \nodepart{lower} $a_6$};
    \node[state with output, right of=s6] (s7) {$\mathbf{s}_7$ \nodepart{lower} $a_7$};
    \node[state with output, right of=s7] (s8) {$\mathbf{s}_8$ \nodepart{lower} $a_8$};
    
    \draw (s0) edge[bend left] node[above]{$a^1_0$} (s2)
          (s0) edge[bend left=50] node[above]{$a^2_0$} (s4)
          (s0) edge[bend left=20] node[below]{$a^3_0$} (s1)
          (s1) edge[bend right=60] node[above]{$a^1_1$} (s3)
          (s1) edge[bend right=60] node[below]{$a^2_1$} (s5)
          (s1) edge[bend right=60] node[above]{$a^3_1$} (s4)
          (s2) edge[bend left=20] node[below]{$a^1_2$} (s3)
          (s2) edge[bend left=50] node[above]{$a^2_2$} (s6)
          (s2) edge[bend left=30] node[above]{$a^3_2$} (s4)
          (s3) edge[bend right=40] node[above]{$a^1_3$} (s8)
          (s3) edge[bend right=50] node[below]{$a^2_3$} (s8)
          (s3) edge[bend right=35] node[above]{$a^3_3$} (s7)
          ;
\end{tikzpicture}
\caption{A system with three different actions, with random action and observation delays.}
\label{fig:delay}
\end{figure*}
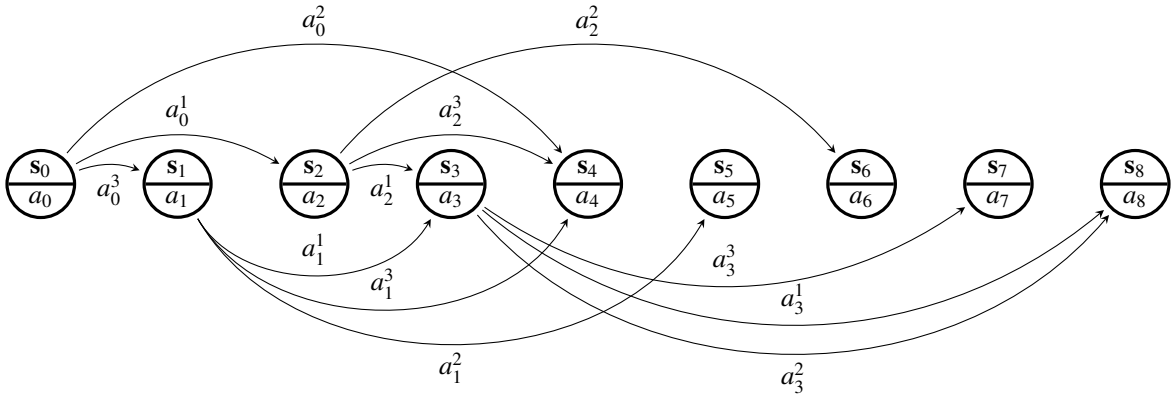

In control systems, constant and random delays in action and observation present significant challenges, particularly in real-time applications such as robotics and remote control. Constant delays, where there is a fixed time lag between action execution or observation and their corresponding feedback, can often be accounted for in system design. However, they still complicate accurate state estimation and timely decision-making \cite{nillson1998real}.

On the other hand, random delays vary unpredictably due to factors like network latency or processing load, introducing greater complexity. These delays can disrupt the synchronization between the system's actions and its perception of the environment, leading to degraded performance or instability. As highlighted in \cite{zhong2014optimal} and \cite{hester2012rtmba}, random delays violate the Markov assumption typically required in control and reinforcement learning algorithms. This necessitates advanced techniques—such as delay compensation or predictive modeling—to maintain system performance. The unpredictable nature of random delays makes them particularly challenging, requiring robust strategies to mitigate their impact on control accuracy and system reliability.

The study in \cite{bouteiller2020reinforcement} addresses action and observation delays in reinforcement learning environments, typical in real-world applications such as remote control operations. The authors introduce the Delay-Correcting Actor-Critic (DCAC) algorithm and the concept of Random-Delay Markov Decision Processes (RDMDPs), which extend traditional Markov Decision Processes (MDP) to incorporate random delays. The DCAC algorithm uses off-policy multi-step value estimation combined with a trajectory resampling technique to transform off-policy trajectories into on-policy ones, reducing the bias introduced by delays and enabling more accurate value estimation. The approach allows off-policy algorithms, such as Soft Actor-Critic, to remain sample-efficient while adapting to variable delays. Experimental results show that DCAC outperforms standard methods like SAC, particularly in challenging scenarios with unpredictable delays.

In specific industrial processes, a common challenge arises from different actions incurring different delays, such that the delays between actions and observations may vary for each action. As illustrated in Figure \ref{fig:delay}, when multiple actions are determined at a single step, the effects of some actions may appear in the observations sooner than others. At each time step $i$, the agent generates a set of actions denoted as $\mathbf{a}_i = \{a^1_i, a^2_i, a^3_i\}$, where $a^1_i$, $a^2_i$, and $a^3_i$ represent individual components of the overall action vector. These actions are determined based on the current state of the system $\mathbf{s}_i$, which encapsulates the system's conditions at time step $i$. The system is characterized by three different actions, each influenced by random action and observation delays, adding complexity to the control dynamics. The agent's actions can be expressed as:
\begin{equation}
\mathbf{a}_i = \pi(\mathbf{s}_{i - d_s}),
\end{equation}

Where $\pi$ is the policy learned by the agent, and $d_s$ represents a random observation delay. Each action component is then defined as:
\begin{equation}
a^1_i = f^1(\mathbf{s}_{i - d_s}), \quad a^2_i = f^2(\mathbf{s}_{i - d_s}), \quad a^3_i = f^3(\mathbf{s}_{i - d_s}),
\end{equation}

In the above equation, $f^1, f^2,$ and $f^3$ are functions representing the mapping from the delayed state $\mathbf{s}_{i - d_s}$ to each action component under the policy $\pi$. Once computed, the system executes these actions with a random action delay $d_a$, resulting in the effective action at the system being:

\begin{equation}
\tilde{a}^j_i = a^j_{i - d_a}, \quad \text{for } j \in \{1, 2, 3\}.
\end{equation}

Here, $\tilde{a}^j_i$ represents the delayed action component that the system perceives at time $i$. The combination of random delays in observation and action introduces significant variability and challenges in maintaining optimal control, requiring the agent to adapt dynamically to the system's stochastic behavior. These delays complicate the credit assignment process, as the agent must wait for the effects of prior actions to manifest in the observations before accurately evaluating their impact. This delay increases the complexity of optimizing such systems, as the temporal misalignment between actions and their observed outcomes can suppress causal relationships.

To address this challenge, this study incorporates a mechanism that accounts for differing delays in action and observation variables. By carefully tracking the temporal impact of each action, the method ensures that credit is assigned only after the effects of the most recent action are reflected in the system's state. This approach enables the optimization algorithm to handle delayed feedback effectively, improving the agent's ability to learn and adapt in environments with asynchronous delays.

\subsection{Contributions}
This study introduces a novel application of the Soft Actor-Critic algorithm for optimizing phosphorus removal in wastewater treatment plants with integrated time delay handling. Key contributions of this work include:

\begin{enumerate}
    \item \textbf{Application of SAC in WWTP Optimization}: This study applies the SAC algorithm within a high-dimensional, time-delayed WWTP simulation, leveraging SAC’s sample efficiency and stability to improve control strategies over conventional Proportional-Integral-Derivative (PID) controllers. This demonstrates the feasibility of SAC as a scalable and adaptive approach for industrial process control in WWTPs.

    \item \textbf{Integration of Time Delays in DRL Training}: Recognizing the challenges posed by time delays in control environments, SAC agents were trained in three distinct scenarios: no delay, constant delay, and random delay. The integration of random delay, in particular, addresses the stochastic nature of time delays in real-world industrial settings, enhancing the robustness and reliability of the trained agents.

    \item \textbf{Use of a Specialized LSTM-based Simulator}: Control policies were trained using a state-of-the-art simulator employing \emph{LSTM} networks, specifically designed to support multi-step simulations with delayed feedback. This simulator enables a realistic evaluation of SAC in a time-delayed environment, allowing the study to accurately assess performance improvements and cost savings.

    \item \textbf{Quantitative Performance Improvements}: The study demonstrates that SAC agents trained with random delay achieved higher reward by \textbf{55}\% and reduced phosphorus emissions by \textbf{36}\% in simulation, compared to traditional PID control. These results underscore SAC's potential to enhance operational efficiency and environmental sustainability in WWTPs.
\end{enumerate}

This study adds to the expanding research on utilizing deep reinforcement learning for industrial process control. It highlights the advantages of incorporating time delay management and leveraging advanced DRL algorithms to address the challenges of complex, multi-variable systems such as wastewater treatment plants.

\section{Methods}
A Soft Actor-Critic algorithm was trained on a custom environment designed with OpenAI's Gym library \cite{brockman2016openai} to optimize phosphorus removal in wastewater treatment plants. While the DCAC algorithm has demonstrated superior performance in handling random delays \cite{bouteiller2020reinforcement}, the Soft Actor-Critic algorithm was selected due to its established stability, ease of implementation, and computational efficiency \cite{haarnoja2018soft}. Additionally, SAC's entropy-regularized framework facilitates robust exploration, which is crucial for managing high-dimensional action spaces in WWTP simulations. The environment utilized an \emph{LSTM} model, which was trained on the plant's dataset to predict the next state of the system based on current state-action pairs, with the predicted observation and calculated reward being returned to the agent. The training was performed using the SAC algorithm implemented in PyTorch \cite{paszke2019pytorch}, and a multi-environment approach was adopted to accelerate the learning process. The Gym asynchronous vectorized environment was used to simulate multiple parallel environments.

\subsection{The Plant and Dataset}
This study focuses on the data from Kolding Central WWTP in Agtrup, Denmark. The time-series dataset for the period of two years was collected through the $\text{Hubgrade}^{\text{TM}}$ Performance Plant system, designed by Krüger/Veolia \cite{hubgrade}. The plant utilizes a combination of biological and chemical methods for phosphorus removal. Chemical phosphorus removal is achieved by adding metal salts at two points within the plant. The first dosing point is located after the primary clarifier and before the biological treatment line, where iron sulfide (JSF) is added. The second dosing point is situated after the biological tank and before the secondary clarifier, where polyaluminum chloride (PAX) is introduced to the stream. Data preprocessing played a crucial role in enhancing model performance. The raw data was initially normalized using the Min-Max technique, scaling the features to a range of 0 to 1. Feature selection was guided by principal component and correlation analysis. Variables of the system that demonstrated the highest correlation with the target variable, Phosphate concentration, were selected as inputs to the model, in conjunction with the target variable itself and the action variables, two points of metal salts dosage. More information about the plant, dataset, and preprocessing is explained in \cite{mohammadi2024deep, mohammadi2024wastewater}.

\subsection{Simulation Environment}
The simulation environment was developed using OpenAI’s Gym library \cite{brockman2016openai}. The \textbf{action space} was defined as vector containing metal salts dosage values:
\begin{equation}
\mathbf{a}_i = \{Q_{v,i}^{\text{JSF}}, Q_{v,i}^{\text{PAX}}\}
\end{equation}

Where $Q_{v,i}^{\text{JSF}}$ and $Q_{v,i}^{\text{PAX}}$ represent the agent's control variables (metal salts dosage) at time step $i$, influencing the system's dynamics. The \textbf{observation space} at time step $i$ was a composite vector defined as:
\begin{equation}
\mathbf{s}_i = \{\mathbf{x}_{e,i}, C_{v,i}^P, \sin(h_i), \cos(h_i), \sin(d_i), \cos(d_i), \sin(m_i), \cos(m_i)\}
\end{equation}

Where $\mathbf{x}_{e,i}$ is the vector of exogenous variables (as explained in \cite{mohammadi2024wastewater, mohammadi2024multi}) at time $i$, $C_{v,i}^P$ is the phosphorus concentration at time $i$, and $h_i$, $d_i$, and $m_i$ represent the hour of the day, day of the week, and month of the year, respectively, encoded as cyclical time features using sine and cosine transformations.

A refined version of the \emph{LSTM} model, enhanced for more accurate multi-step simulations as described in \cite{mohammadi2024deep, mohammadi2024improved, mohammadi2024multi}, was employed as a predictor to forecast the system’s next state. After each prediction, the observation and reward were returned to the agent. Utilizing the \emph{LSTM} predictor, the core environment provided only the predicted observation for the next state based on the agent’s actions.

Inspired by the approach in \cite{bouteiller2020reinforcement}, custom wrappers were created to handle the complexity introduced by random action and observation delays. These wrappers expanded the observation space by incorporating additional variables, such as an action buffer, set to the maximum of the observation and action delays, and separate delay indicators for both action and observation delays. This design ensured the agent could adapt effectively to the asynchronous nature of the environment while maintaining robustness in optimizing the control process.

\subsection{Reward Function}
The reward function was designed to train a Soft Actor-Critic algorithm for optimizing phosphorus removal in a wastewater treatment plant. The function evaluates the cost-effectiveness of controlling phosphorus concentration while minimizing financial penalties, taxes, and action costs.

\subsubsection{Reward Calculation Overview}
The reward function is structured to minimize the total cost associated with phosphorus control, including penalties for deviations from a target phosphorus concentration, taxation based on the phosphorus load, and costs for applying chemicals (JSF and PAX). The overall reward is calculated as follows:
\begin{equation}
r = - C_{\text{total}} \cdot (1 + P_{\text{coef}})
\label{eq:reward-first}
\end{equation}

Where $r$ is the reward, $C_{\text{total}}$ includes chemical costs and tax, and $P_{\text{coef}}$ is a coefficient calculated by the penalty function based on the deviation from the target phosphorus concentration.

\subsubsection{Total Cost Calculation}
\paragraph{Action Costs}
The action costs include the expenses for the chemical precipitants JSF and PAX, calculated as follows:
\begin{subequations}
\begin{align}
C_{\text{JSF}} &= Pr_{\text{JSF}} \cdot Q_v^{\text{JSF}} \cdot \frac{t_{\text{dose}}}{60} 
\\
C_{\text{PAX}} &= Pr_{\text{PAX}} \cdot Q_v^{\text{PAX}} \cdot \frac{t_{\text{dose}}}{60}
\end{align}
\end{subequations} 

Here, $C_{\text{JSF}}$ and $C_{\text{PAX}}$ represent the costs of the metal salts, while $Pr_{\text{JSF}}$ and $Pr_{\text{PAX}}$ denote their respective prices in currency per liter, which are considered to be 0.20 and 3.54 Danish kroner per liter based on the Danish market. Additionally, $Q_v^{\text{JSF}}$ and $Q_v^{\text{PAX}}$ are the flow rates of the metal salts in liters per hour, and $t_{\text{dose}}$ is the dosing duration in minutes, corresponding to the sampling frequency in the dataset.

\paragraph{Tax Costs}
The tax is calculated based on the phosphorus load in the wastewater. The phosphorus mass, $M_P$, in kilograms is determined using:

\begin{equation} M_P = C_v^P \cdot Q_v^w \cdot \frac{t_{\text{dose}}}{60000} \end{equation}

Where $M_P$ represents the mass of phosphorus in kilograms ($kg$), $C_v^P$ is the phosphorus concentration in milligrams per liter ($mg/L$), and $Q$ is the flow rate of wastewater in cubic meters per hour ($m^3/h$). Given that $C_v^P$, $Q_v^w$, and $t_{\text{dose}}$ are expressed in $mg/L$, $m^3/h$, and minutes, dividing by $60000$ ensures that the result is in kilograms. Using this calculated phosphorus mass, the tax cost can then be determined as follows:
\begin{equation}
T = T_{\text{rate}} \cdot M_P
\end{equation}

Here, $T$ and $T_{\text{rate}}$ are the amount of tax paid by the plant and the green tax rate in currency per kilogram of phosphorus (183.64 Danish Krones per kilogram in Denmark). Finally, the total cost of each dosing step is calculated as follows:
\begin{equation}
    C_{\text{total}} = C_{\text{JSF}} + C_{\text{PAX}} + T
\label{eq:total-cost}
\end{equation}

\subsubsection{Penalty Coefficient Calculation}
The penalty function measures the deviation from an ideal phosphorus concentration target and varies depending on the approach used (Linear or Non-linear):

\paragraph{Linear Penalty}
For the Linear approach, the penalty is calculated as:
\begin{equation}
P_{\text{coef}} =
\begin{cases}
0, & \text{if } 0 < x \leq x_{\text{ideal}} \\
-100 \cdot T, & \text{otherwise}
\end{cases}
\end{equation}

Where $x$ represents the current phosphorus concentration and $x_{\text{ideal}}$ is the ideal target level for the phosphorus concentration, which should be under the emission limit according to the regulations.

\paragraph{Non-linear Penalty}
For the Non-linear approach, the penalty is computed using a custom exponential function:
\begin{equation}
P_{\text{coef}}(x) = a \cdot e^{z \cdot x + c} + d
\label{eq:nonlinear-penalty}
\end{equation}

Where $a$, $z$, $c$, and $d$ are parameters controlling the shape and scale of the exponential penalty, and $x$ is the phosphorus concentration. The parameters were selected to make the penalty relatively mild when the concentration is below the ideal target level and near zero. Still, the penalty becomes steeper and more severe as the concentration exceeds the limit. 
A non-linear penalty was selected for this study because it increases progressively with concentration levels, allowing for more significant differentiation between minor and major violations.

\subsubsection{Summary of the Reward Function}
The overall reward function thus integrates costs and penalties to guide the SAC algorithm towards actions that minimize deviations from optimal phosphorus levels while keeping treatment costs low. According to the equations \ref{eq:reward-first}, \ref{eq:total-cost}, and \ref{eq:nonlinear-penalty}, the structure of this reward function can be defined as:
\begin{equation}
r_t(x) = - (C_{\text{JSF},t} + C_{\text{PAX},t} + T_t) \cdot (1 + a \cdot e^{z \cdot x_t + c} + d) 
\end{equation}

This formulation ensures that higher costs or deviations from the target concentration result in lower rewards, thereby effectively incentivizing the agent to optimize the phosphorus removal process.

\subsection{Soft Actor-Critic}
Actor-critic methods are a class of reinforcement learning algorithms that involve two primary components: an actor (policy) network and a critic (value) network. The actor network is responsible for selecting actions given the current state, while the critic network evaluates the quality of those actions by estimating the value function. Soft Actor-Critic is an actor-critic algorithm incorporating entropy maximization to encourage exploration \cite{haarnoja2018soft}.

\subsubsection{Key Components and Equations in SAC}
\paragraph{Policy Objective with Entropy Regularization} 
The SAC algorithm’s policy objective maximizes the expected return while encouraging exploration by maximizing the entropy of the policy. This is formulated as \cite{haarnoja2018soft}:
\begin{equation}
    J_\pi(\phi) = \mathbb{E}_{\mathbf{s}_t \sim \mathcal{D}, \mathbf{a}_t \sim \pi_\phi} \left[ \alpha \mathcal{H}(\pi(\cdot|\mathbf{s}_t)) - Q_{\theta}(\mathbf{s}_t, \mathbf{a}_t) \right]
\end{equation}

In this equation, $\alpha$ represents the temperature parameter, which balances the trade-off between exploration and exploitation. Higher values of $\alpha$ increase the weight of the entropy term, encouraging more exploration. The entropy term, $\mathcal{H}(\pi_\phi(\cdot|\mathbf{s}_t))$, quantifies the randomness in the action distribution at state $\mathbf{s}_t$, promoting a stochastic policy that explores diverse actions. Additionally, $Q_{\theta}(\mathbf{s}_t, \mathbf{a}_t)$ denotes the soft Q-value function, parameterized by $\theta$, which estimates the expected return for taking action $\mathbf{a}_t$ in state $\mathbf{s}_t$. Finally, $\pi_\phi$ refers to the specific learned policy with parameters $\phi$, while $\pi$ generally represents the policy distribution over actions. This objective encourages $\pi_\phi$ to explore uncertain or high-reward regions, facilitating robust performance by promoting varied behavior across different states.  The policy $\pi_\phi$, parameterized by $\phi$, typically uses neural network parameters for function approximation.

\paragraph{Q-Value Objective}
The Q-value function objective in SAC minimizes the Bellman error, which measures the difference between the predicted and target Q-values. This objective can be expressed as \cite{haarnoja2018soft}:
\begin{equation}
   J_Q(\theta) = \mathbb{E}_{\mathbf{s}_t, \mathbf{a}_t \sim \mathcal{D}} \left[ \frac{1}{2} \left( Q_{\theta}(\mathbf{s}_t, \mathbf{a}_t) - \left( r(\mathbf{s}_t, \mathbf{a}_t) + \gamma \mathbb{E}_{\mathbf{s}_{t+1} \sim p} \left[ V(\mathbf{s}_{t+1}) \right] \right) \right)^2 \right]
\end{equation}

Here, $r(\mathbf{s}_t, \mathbf{a}_t)$ is the reward for executing action $\mathbf{a}_t$ in state $\mathbf{s}_t$, and $\gamma$ is the discount factor, which controls the weighting of future rewards relative to immediate rewards. The target Q-value incorporates the soft value of the next state $V(\mathbf{s}_{t+1})$, which is computed as \cite{haarnoja2018soft}:
\begin{equation}
    V(\mathbf{s}_{t+1}) = \mathbb{E}_{\mathbf{a}_{t+1} \sim \pi_\phi} \left[ Q_\theta(\mathbf{s}_{t+1}, \mathbf{a}_{t+1}) - \alpha \log \pi_\phi(\mathbf{a}_{t+1} | \mathbf{s}_{t+1}) \right]
\end{equation}

The soft value $V(\mathbf{s}_{t+1})$ combines the expected Q-value for the next state-action pair with an entropy term $\alpha \log \pi_\phi(\mathbf{a}_{t+1} | \mathbf{s}_{t+1})$, which promotes exploratory behavior by increasing the value of uncertain or high-entropy actions. These objectives balance exploration and exploitation, improving sample efficiency and stability in the training process.

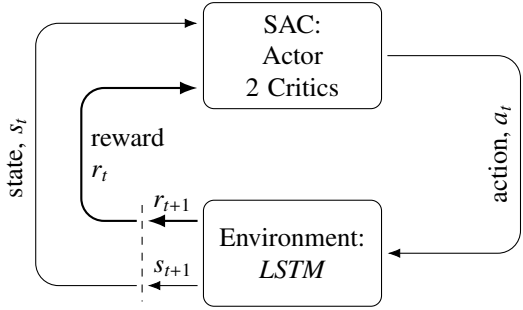
\begin{figure}
\centering
\begin{tikzpicture}[auto, node distance = 4cm]
    \node[frame] (agent) {SAC:\\Actor\\2 Critics};
    \node[frame, below=1.2cm of agent] (environment) {Environment: \emph{LSTM}};
    
    \draw[line-rounded] (agent) -- ++ (3.0, 0) |- (environment)
    node[above left, pos=0.1, align=center, rotate=90] {action, $a_t$};
    \coordinate[left=8mm of environment] (P);
    \draw[thin,dashed, -] (P|-environment.north) -- (P|-environment.south);
    \draw[line-rounded] (environment.200) -- (P |- environment.200)
    node[midway,above]{$s_{t+1}$};
    \draw[line-rounded,thick] (environment.160) -- (P |- environment.160)
    node[midway,above]{$r_{t+1}$};
    \draw[line-rounded] (P |- environment.200) -- ++ (-1.4,0) |- (agent.160)
    node[above right, pos=0.15, align=center, rotate=90] {state, $s_t$};
    \draw[line-rounded,thick] (P |- environment.160) -- ++ (-0.8,0) |- (agent.200)
    node[right,pos=0.25,align=left] {reward\\ $r_t$};
\end{tikzpicture}
\caption{The process of training Soft Actor-Critic policy on the simulation environment}
\label{fig:drl_delay}
\end{figure}

\subsection{Training Procedure}
The SAC agent was set up with one actor and two critics for training, with each critic independently estimating the expected return for the current policy. During each training step, the agent applies a set of actions and receives the next state and reward from the environment. The two critics provide separate Q-value estimates to reduce overestimation bias by taking the minimum value during policy updates, enhancing training stability. The observation (state) space varies depending on whether delays are accounted for. In scenarios with delays, the state space includes delay-related variables. Figure \ref{fig:drl_delay} illustrates the overall training procedure, highlighting the inclusion of delay variables where applicable. To leverage the parallelism offered by the asynchronous vector environment, we configured 16 separate environments, and each initialized with different settings to ensure a diverse range of initial observations and episode lengths. This diversity was intended to improve the robustness of the training process by exposing the SAC algorithm to varied starting conditions and episode dynamics. The 16 environments were divided into four experimental setups, with four environments assigned to each setup:

\begin{itemize}
    \item \textbf{E1 - Constant-Length, Consecutive Episodes:} This baseline experiment used episodes of equal length, sampled sequentially from the time-series data. It aimed to evaluate the model's performance under a stable and predictable training regime.
    \item \textbf{E2 - Random-Length, Consecutive Episodes:} Episodes followed a sequential order but varied in length, introducing unpredictability in the steps per episode. This experiment aimed to simulate real-world scenarios with fluctuating sequence steps.
    \item \textbf{E3 - Random Start, Constant-Length Episodes:} Episodes of constant length were initialized at random points within the dataset. This setup tested the algorithm's ability to learn from unsequentially organized data, introducing randomness in the initial state while keeping the number of steps consistent.
    \item \textbf{E4 - Random Episodes with Random Lengths:} Both the starting points and lengths of episodes were randomized. This scenario challenges the algorithm's adaptability to an environment where the initial state and the number of steps are highly variable.
\end{itemize}

\begin{algorithm}
\caption{Soft Actor-Critic Training with Multiple Environments and Delay Handling}
\label{alg:sac_training}
\begin{algorithmic}[1]
\State \textbf{Initialize} policy $\pi_\phi$, Q-value networks $Q_{\theta_1}$, $Q_{\theta_2}$, target Q-networks $\hat{Q}_{\theta_1}$, $\hat{Q}_{\theta_2}$, replay buffer $\mathcal{D}$, and 16 environments
\State \textbf{Divide} environments into four setups: E1, E2, E3, and E4 
\State \textbf{Initialize} delay settings: \textit{No}, \textit{Constant}, or \textit{Random Delay}
\If{\textit{Constant} or \textit{Random Delay}} 
\State \textbf{Initialize} the range of action delay with ($\kappa_{\text{min}}, \kappa_{\text{max}}$)
\State and observation delay with ($\omega_{\text{min}}, \omega_{\text{max}}$)
\State \textbf{Initialize} action buffer $\mathcal{B}_a$ with the size $(\kappa_{\text{max}} + \omega_{\text{max}})$
\EndIf
\For{each training step}
    \For{each environment $e_i$ in parallel}
        \State \textbf{Observe} state $s_t^{(e_i)}$
        \State \textbf{Select} action $a_t^{(e_i)} \sim \pi_\phi(a_t^{(e_i)} | s_t^{(e_i)})$
        \State \textbf{Apply} $a_t^{(e_i)}$, \textbf{observe} $s_{t+1}^{(e_i)}$, $r_t^{(e_i)}$, delays if applicable
        \If{\textit{Constant Delay}}
            \State $\kappa_t \gets \kappa_{\text{max}}$ and $\omega_t  \gets \omega_{\text{max}}$
        \ElsIf{\textit{Random Delay}}
            \State $\kappa_t \sim \text{DiscreteUniform}(\alpha_{\text{min}}, \kappa_{\text{max}})$
            \State $\omega_t  \sim \text{DiscreteUniform}(\omega_{\text{min}}, \omega_{\text{max}})$
        \Else 
            \State $\kappa_t \gets 0$ and $\omega_t  \gets 0$
        \EndIf
        \If{\textit{Constant} or \textit{Random Delay}}
        \State \textbf{Update} $\mathcal{B}_a$ with $a_t$
        \State $\tau_t \gets \left[\kappa_t, \omega_t, \mathcal{B}_a \right]$
        \State $s_t \gets \left[s_t, \tau_t\right]$
        \EndIf
        \State \textbf{Store} transition $\mathcal{D} \leftarrow (s_t^{(e_i)}, a_t^{(e_i)}, r_t^{(e_i)}, s_{t+1}^{(e_i)})$
    \EndFor
\If{ready to update}
    \For{each gradient step}
        \State \textbf{Sample} $N$ transitions $\{ (s_i, a_i, r_i, s_{i+1}) \}_{i=1}^N \sim \mathcal{D}$
        
        \State $y_i \gets r_i + \gamma \min_{j=1,2} \hat{Q}_{\theta_j}(s_{i+1}, \pi_\phi(s_{i+1})) -\alpha \log \pi_\phi(a_{i+1} | s_{i+1})$
        
        \State $L(\theta_j) \gets \frac{1}{N} \sum_{i} \left( Q_{\theta_j}(s_i, a_i) - y_i \right)^2 \text{ for } j = 1,2$
        
        \State $\nabla_\phi J(\phi) \gets \frac{1}{N} \sum_{i} \left( \nabla_\phi \log \pi_\phi(a_i | s_i) \right. \times \left[ \min_{j=1,2} Q_{\theta_j}(s_i, a_i) - \alpha \log \pi_\phi(a_i | s_i) \right] )$
        
        \State $\hat{\theta}_j \gets \tau \theta_j + (1 - \tau) \hat{\theta}_j \text{ for } j = 1,2$
    \EndFor
\EndIf

        
        
        
        
    
\EndFor
\end{algorithmic}
\end{algorithm}

\subsubsection{Incorporation of Delays}
Three different policy learning scenarios were studied: \textit{No Delay}, \textit{Constant Delay}, and \textit{Random Delay}. In the \textit{No Delay} scenario, the observation space included only the predicted state of the system returned by the environment. In contrast, the \textit{Constant Delay} and \textit{Random Delay} scenarios incorporated additional information in the observation space, such as the action buffer ($\mathcal{B}_a$), observation delay ($\omega$), and action delay ($\kappa$). For the \textit{Constant Delay} scenario, the delay at each step for both action and observation was set to a constant value equal to the maximum of the delay ranges. In the \textit{Random Delay} scenario, the action and observation delays were sampled randomly from their respective ranges at each step, introducing variability in the timing of actions and observations. This setup was designed to evaluate the algorithm's ability to learn robust policies under different levels of temporal uncertainty.

A concise overview of the SAC training procedure, including the integration of delay handling, is presented in Algorithm \ref{alg:sac_training}.

\subsection{Comparison to the Current Controller}
\label{sec:pid-method}
In this study, the real-time PID controller currently operating in the wastewater treatment plant, which could dynamically produce actions in response to system states, was not accessible. Instead, historical data were utilized to replicate the past PID controller's actions for similar system states.

Specifically, for a given input sequence, we assumed that the PID controller would make the same decisions as it did historically for the corresponding state. The actions taken by the PID controller at time $t$ were retrieved from the dataset, where they were initially logged as the controller's response to the system state $\mathbf{s}_t$. This array of action variables effectively represents the PID's decision-making for the sequence in question. Essentially, the actions produced by the PID controller at time $t$ are assumed to be an array of the system's past actions at that time:
\begin{equation}
    \mathbf{a}_{\text{PID},t} = \{Q_{v,t}^{\text{JSF}}, Q_{v,t}^{\text{PAX}}\}
\end{equation}

In the above equation,     $\mathbf{a}_{\text{PID},t}$, $Q_{v,t}^{\text{JSF}}$, and $Q_{v,t}^{\text{PAX}}$ represent the actions produced by the PID controller, the flow of the iron salts dosage, and the flow of the PAX dosage at time $t$.

The same input sequence was provided to all controllers to compare the performance of the PID controller and the Soft Actor-Critic agents. SAC agents generated actions based on the observed system state, while the PID actions were directly taken from the dataset. The actions of both the PID controller and SAC agents were then sent to the custom simulator, which produced the next state of the system, enabling a side-by-side evaluation of their performance.

\subsection{Software and Hardware}
All of the tests for the simulation environment are implemented in programming language \emph{Python} by using the \emph{Gym} \cite{brockman2016openai} and \emph{PyTorch} \cite{NEURIPS2019_9015} libraries. The AI Cloud service from Aalborg University is used for GPU-based computations. The available compute nodes are each equipped with 2 × 24-core \emph{Intel Xeon} CPUs, 1.5 TB of system RAM, and 16 \emph{NVIDIA Tesla V100} GPUs with 32 GB of RAM each, all connected via \emph{NVIDIA NVLink}. One compute node consisting of one GPU and 32 CPU cores was used to train each experiment.

\section{Results}
\subsection{Policy Learning}
\begin{figure}
\centering
\includegraphics{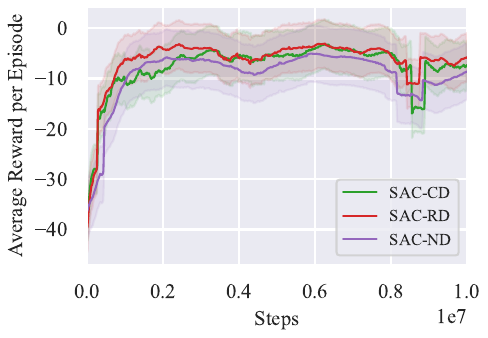}
\caption{Average cumulative rewards per episode for SAC with different delay and Multi-experiment with \textbf{non-linear} reward function.}
\label{fig:non_linear_rw}
\end{figure}

The training of the Soft Actor-Critic algorithm was conducted across three distinct scenarios: No Delay (ND), Constant Delay (CD), and Random Delay (RD). Each scenario aimed to evaluate the impact of different temporal dynamics on the performance of the control policies. Figure \ref{fig:non_linear_rw} illustrates the episodic average reward per step versus the number of total environment steps the agent takes. The results indicate that delay-aware models (CD and RD) significantly outperformed the No Delay scenario. This performance gap emerged after approximately 10 million training steps, where the agents accounting for delays began to accumulate higher rewards.

The Random Delay scenario demonstrated the highest cumulative rewards across training, indicating that the SAC agent was better able to handle the variability in delays between actions and their effects. This result is particularly noteworthy because real-world wastewater treatment plants often experience fluctuating delays due to various factors, including mechanical issues and sensor delays. The ability to manage such uncertainties is crucial for maintaining optimal plant operation.

The agent struggled to perform in the No Delay scenario, accumulating lower rewards throughout training. This outcome can be attributed to the model's inability to account for the inherent delays in WWTP operations, leading to suboptimal action choices and less efficient phosphorus removal.

The reward structure was designed to balance multiple factors, including minimizing chemical costs and meeting phosphorus concentration targets. In the CD and RD scenarios, the SAC algorithm learned policies that more effectively struck this balance, using fewer resources while maintaining regulatory compliance. The results underscore the importance of incorporating delay mechanisms into reinforcement learning models, especially in industrial systems where time lags can significantly impact process performance.
\subsection{Comparison with the Existing Control Policy}
\begin{figure}
    \centering
    \includegraphics[]{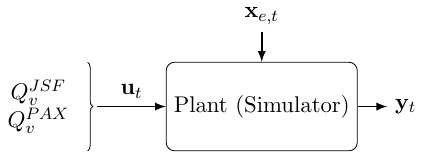}
    \caption{The inputs and outputs of the phosphorus removal process in WWTP. Where $\mathbf{u}_t$, $\mathbf{x}_{e,t}$, and $\mathbf{y}_t$ represent the control variables, exogenous variables, and target variables.}
    \label{fig:input_output}
\end{figure}

To further evaluate the effectiveness of the learned policies, the best-performing SAC agents (ND, CD, RD) were compared against the PID controller currently used at the WWTP. The PID controller, a common choice in industrial control systems, operates reactively, relying on historical data to address deviations from a setpoint without forecasting future changes. The comparison methodology, detailed in Section \ref{sec:pid-method}, involves replicating the actions taken by the PID controller for the same initial state in the past. For a given historical initial state, the performance of the trained agents was analyzed in the simulation environment. As illustrated in Figure \ref{fig:input_output}, the environment predicts the output or target variables at each simulation step based on the control inputs generated by the agents and the exogenous variables from the dataset corresponding to that time step.
\begin{figure*}[h]
\centering
\includegraphics{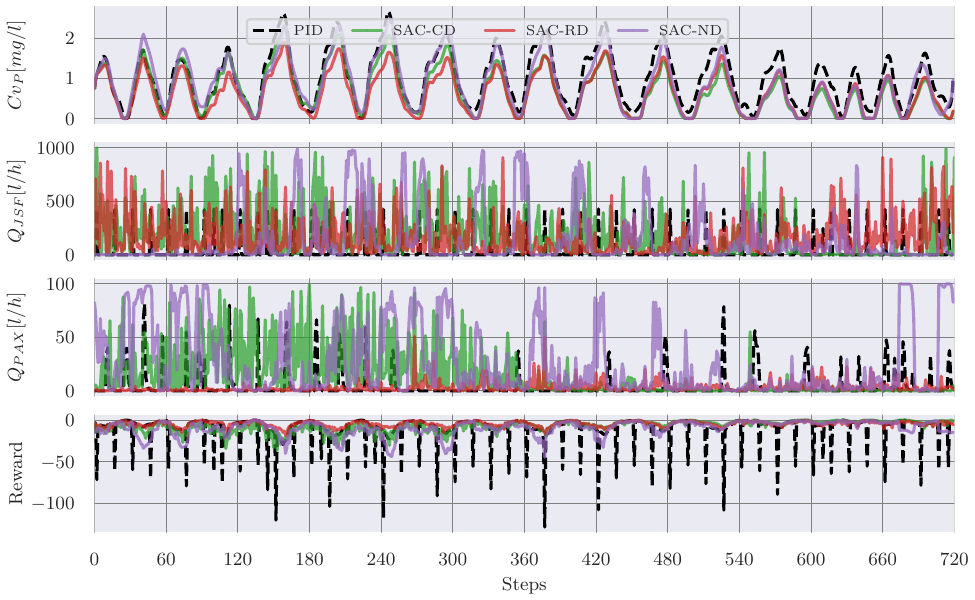}
\caption{The comparison of the existing PID control and learned SAC policies for a point of the wastewater treatment dataset on September 15th, 2022.}
\label{fig:eval_plot}
\end{figure*}

Figure \ref{fig:eval_plot} presents the performance comparison, focusing on two critical aspects: the concentration of phosphate in the effluent and the control actions involving the dosage of metal salts. The SAC agents, particularly the one trained under random delay conditions, demonstrated superior control over the phosphate levels. The SAC policies maintained the phosphate concentration closer to the target value, with fewer fluctuations than the PID controller. This ability to keep the system stable is crucial, as even minor deviations from the target phosphorus level can result in penalties or increased operational costs.
\begin{table*}
\caption{Evaluation Metrics for the current PID control and SAC Agents}
\label{tab:eval_metrics}
\resizebox{\textwidth}{!}{%
\begin{tabular}{lcccccccc}
\toprule
Agent & Tot. Reward & Avg. Reward & Avg. Target & Tot. $C_{\text{JSF}}$ & Tot. $C_{\text{PAX}}$ & Tot. Tax & Tot. Costs & Target Dev. \% \\
\midrule
PID & -7275.23 & -10.10 & 0.98 & \textbf{149.17} & 354.27 & 2421.25 & 2924.69 & 24.17 \\
SAC-CD & -4971.40 & -6.77 & 0.69 & 886.61 & 1204.12 & 1780.73 & 3871.47 & 9.81 \\
SAC-RD & \textbf{-3293.68} & \textbf{-4.49} & \textbf{0.62} & 806.57 & \textbf{267.3} & \textbf{1578.73} & \textbf{2652.6} & \textbf{5.59} \\
SAC-ND & -6940.70 & -9.64 & 0.85 & 657.00 & 2192.79 & 2126.94 & 4976.74 & 19.17 \\
\bottomrule
\end{tabular}
}
\end{table*}

Regarding resource efficiency, the SAC agents also outperformed the PID controller. The learned policies more effectively optimized the dosage of metal salts, reducing the overall consumption of chemicals while maintaining adequate phosphorus removal. This improvement translates into cost savings for the plant, as chemicals like iron chloride and polyaluminum chloride represent a significant portion of the operational expenses in phosphorus removal.

During the evaluation, step-by-step rewards were calculated for both control strategies. The SAC agents consistently achieved higher rewards, indicating that their actions were more cost-effective and better aligned with the WWTP's operational goals. The SAC agents' ability to adapt to fluctuating conditions, particularly in the RD scenario, highlights the potential of deep reinforcement learning (DRL) to improve process control in complex, dynamic environments like wastewater treatment plants.

\section{Discussion}
The results of this study underscore the critical importance of accounting for time delays in optimizing control policies for wastewater treatment plants. Due to the systems' complexity, time delays between actions and their observed effects are inevitable in industrial processes like WWTPs. Ignoring these delays, as seen in the no delay scenario, leads to suboptimal control performance, as the agent cannot make informed decisions based on delayed feedback.

Figure \ref{fig:non_linear_rw} emphasizes that incorporating constant and random delays into the SAC training process improved the agent's ability to optimize phosphorus removal. The RD scenario, in particular, stands out as the most effective, demonstrating the highest rewards and the most efficient resource usage. This suggests that the variability in delay times, a common occurrence in real-world systems, is best handled by the agents adapting to changing conditions rather than assuming a fixed delay structure.

\subsection{Effect of Delay Incorporations}
Time delays are inherent in wastewater treatment processes due to the plant's complex system dynamics and slow response times. When these delays are not explicitly accounted for in the reinforcement learning framework, the agent's ability to make informed and effective decisions is compromised. 

As illustrated in Figures \ref{fig:non_linear_rw} and \ref{fig:eval_plot}, failing to account for these delays caused the reinforcement learning agent to make decisions that were poorly synchronized with the system's true state. This misalignment resulted in suboptimal action selection, as the agent struggled to anticipate the delayed effects of its actions on system responses. Consequently, the system experienced inefficiencies in phosphorus removal, characterized by greater fluctuations in phosphate concentrations and increased operational costs.

For example, in the no delay scenario, the agent's actions were based solely on the immediate state observations, disregarding the plant's temporal dynamics. This resulted in reactive rather than anticipatory behavior, where the agent failed to compensate adequately for the lagged effects of previous actions. As a result, the control policy struggled to maintain stable phosphorus levels within regulatory limits, demonstrating the critical importance of incorporating delay mechanisms in reinforcement learning models for dynamic and time-sensitive systems like wastewater treatment plants.

\subsection{Comparison to the Current Controller}
Figure \ref{fig:eval_plot} and Table \ref{tab:eval_metrics} highlight a trade-off between costs and target deviations among the control strategies. While the PID controller sometimes achieves lower total costs, such as \textbf{41}\% lower than SAC-ND and \textbf{26}\% lower than SAC-CD, it shows a significantly higher target deviation, with \textbf{77}\% more deviation than SAC-RD and \textbf{59}\% more than SAC-CD. This deviation indicates a greater risk of exceeding the phosphorus effluent limits set by regulations.

Meeting effluent target limits is the primary objective in phosphorus removal, as non-compliance can result in penalties and environmental harm. Cost optimization becomes relevant only when regulatory compliance is guaranteed. The SAC-RD agent demonstrates a clear advantage by achieving the lowest target deviation (\textbf{5.6}\%), which is \textbf{76.9\% lower} than PID, while simultaneously reducing total costs by \textbf{9}\% compared to PID. SAC-CD also reduces the deviation by \textbf{59}\% but at a higher cost.

These findings underscore the limitations of the PID controller in balancing regulatory compliance and cost efficiency. SAC agents, particularly SAC-RD, provide superior control performance by maintaining phosphorus levels within regulatory limits and optimizing costs, making them better suited for real-world wastewater treatment applications.

\subsection{Practical Implications}
The findings of this study have significant practical implications for the control of WWTPs and similar industrial processes. The ability to optimize control strategies in the presence of delays can lead to more efficient operations, reduced costs, and improved regulatory compliance. The results suggest that WWTPs could benefit from transitioning from traditional PID controllers to DRL-based systems like SAC, which are better equipped to handle the complexities and uncertainties inherent in wastewater treatment.

Additionally, the study highlights the potential for DRL to be integrated with existing control systems. While the SAC algorithm outperformed the PID controller in this study, there may be scenarios where a hybrid approach could be beneficial. For example, the PID controller could handle routine operations while the DRL system manages more complex decision-making tasks that require anticipation of future states.

\section{Conclusions}
This study presented a comprehensive framework for optimizing phosphorus removal in wastewater treatment plants using the Soft Actor-Critic algorithm, emphasizing handling time delays in system dynamics. Incorporating constant and random delays into the reinforcement learning framework proved essential for improving the robustness and performance of control policies in dynamic and uncertain environments. By addressing these delays, the SAC agents demonstrated significant improvements in maintaining phosphorus levels within regulatory limits while optimizing operational costs.

The results highlight the effectiveness of delay-aware models, particularly under random delay scenarios, which align more closely with the stochastic nature of real-world systems. The SAC-RD agent achieved the highest performance, reducing target deviations by \textbf{77}\%, lowering total costs by \textbf{9}\%, and decreasing phosphorus emissions by \textbf{36}\% compared to traditional PID controllers. These findings underscore the importance of flexible and adaptive control strategies, as well as the ability of advanced RL algorithms to anticipate system responses and dynamically adjust control actions.

The integration of \emph{LSTM} simulators and custom delay-handling mechanisms enhanced the RL framework, enabling realistic training and evaluation of control policies. This study provides a foundation for applying reinforcement learning in industrial processes, showcasing its ability to adapt to complex, multi-variable environments like WWTPs.

Future research could explore the integration of multi-objective optimization to more effectively balance environmental and economic objectives. Additionally, combining RL with model-based control approaches or hybrid systems could enhance scalability and robustness, paving the way for broader applications in industrial process control.

By addressing the challenges of delayed feedback and optimizing chemical usage, this study contributes to the advancement of sustainable wastewater treatment practices, offering a path toward more efficient and cost-effective solutions for environmental management.

\section{Acknowledgements}
The RecaP project has received funding from the European Union’s Horizon 2020 research and innovation programme under the Marie Skłodowska-Curie grant agreement No 956454. Disclaimer: This publication reflects only the author's view; the Research Executive Agency of the European Union is not responsible for any use that may be made of this information.

\bibliographystyle{elsarticle-num} 
\bibliography{cas-refs}

\end{document}